\documentclass{article}

\usepackage{arxiv}

\usepackage[utf8]{inputenc} % allow utf-8 input
\usepackage[T1]{fontenc}    % use 8-bit T1 fonts
\usepackage{hyperref}       % hyperlinks
\usepackage{url}            % simple URL typesetting
\usepackage{booktabs}       % professional-quality tables
\usepackage{amsfonts}       % blackboard math symbols
\usepackage{nicefrac}       % compact symbols for 1/2, etc.
\usepackage{microtype}      % microtypography
\usepackage{cleveref}       % smart cross-referencing
\usepackage{lipsum}         % Can be removed after putting your text content
\usepackage{graphicx}
\usepackage[numbers]{natbib}
\usepackage{float}
\usepackage{doi}
\usepackage{soul} % for the command \hl
\usepackage{color} % often needed for soul
\usepackage{xcolor}

\newcommand{\orcidicon}{\IfFileExists{orcid.pdf}{\includegraphics[scale=0.06]{orcid.pdf}}{}}

\title{Graphs RAG at Scale: Beyond Retrieval-Augmented Generation with Labeled Property Graphs and Resource Description Framework for Complex and Unknown Search Spaces}

\newif\ifuniqueAffiliation
\uniqueAffiliationtrue

\ifuniqueAffiliation % Standard variant of author block
\author{
	Manie Tadayon\\ AI Scientist, Capital Group \\
	\texttt{manie.tadayon@capgroup.com}
	\And
	Mayank Gupta\\ AI Scientist, Capital Group \\
	\texttt{mayank.gupta@capgroup.com}
}
\renewcommand{\shorttitle}{}

\hypersetup{
pdftitle={Graphs Unleashed: Retrieval-Augmented Generation with Labeled Property Graphs and RDF for Complex and Unknown Search Spaces},
pdfsubject={q-bio.NC, q-bio.QM},
pdfauthor={Manie Tadayon, Mayank Gupta},
pdfkeywords={Graph, Graph RAG, RAG, Retrieval-Augmented Generation, Amazon Neptune, LPG, RDF, Knowledge Graph, OpenCypher},
}

\begin{document}
\maketitle

\begin{abstract}
	Recent advances in Retrieval-Augmented Generation (RAG) have revolutionized knowledge-intensive tasks, yet traditional RAG methods struggle when the search space is unknown or when documents are semi-structured or structured. We introduce a novel end-to-end Graph RAG framework that leverages both Labeled Property Graph (LPG) and Resource Description Framework (RDF) architectures to overcome these limitations. Our approach enables dynamic document retrieval without the need to pre-specify the number of documents and eliminates inefficient reranking. We propose an innovative method for converting documents into RDF triplets using JSON key-value pairs, facilitating seamless integration of semi-structured data. Additionally, we present a text to Cypher framework for LPG, achieving over 90\% accuracy in real-time translation of text queries to Cypher, enabling fast and reliable query generation suitable for online applications. Our empirical evaluation demonstrates that Graph RAG significantly outperforms traditional embedding-based RAG in accuracy, response quality, and reasoning, especially for complex, semi-structured tasks. These findings establish Graph RAG as a transformative solution for next-generation retrieval-augmented systems.
\end{abstract}

% keywords can be removed

\section{Introduction}
Large Language Models (LLMs) have rapidly transformed the landscape of machine learning and artificial intelligence, enabling breakthroughs in natural language understanding, question answering, summarization, and creative text generation. Their remarkable ability to synthesize information and generate coherent, contextually relevant responses has unlocked new possibilities across domains such as healthcare, finance, law and technology \cite{chen2024survey,chkirbene2024large,denecke2024potential}. However, despite their impressive capabilities, LLMs are fundamentally limited by the static nature of their training data, often leading to outdated knowledge and, more critically, the phenomenon of hallucination—where the model generates plausible-sounding but factually incorrect or unverifiable information. Hence it is crucial to augment LLMs with dynamic, reliable sources of information to improve accuracy and reduce hallucinations.

One approach to mitigate the limitations of static LLMs is continual pretraining and supervised fine-tuning, where the model is further trained on new or specialized data to incorporate recent knowledge or adapt to particular tasks. While this strategy can enhance model relevance and performance in targeted domains, it is often prohibitively time-consuming and computationally expensive. Continual fine-tuning requires access to large-scale, high-quality datasets and significant hardware resources, Moreover, fine-tuning may risk catastrophic forgetting of previously acquired knowledge and can introduce new biases or inconsistencies. These challenges make continual  fine-tuning impractical for many real-world applications that demand timely, cost-effective, and reliable access to up-to-date information.

To address these challenges, Retrieval-Augmented Generation (RAG) has emerged as a powerful paradigm, combining the generative prowess of LLMs with dynamic access to external knowledge sources \cite{lewis2020retrieval,jiang2023active,fan2024survey,arslan2024survey}. By retrieving relevant documents or facts at inference time, RAG systems can ground their responses in up-to-date, authoritative information, significantly reducing hallucinations and improving factual accuracy. This approach has proven invaluable for applications requiring real-time knowledge, such as customer support, legal research, and scientific discovery.

Yet, traditional RAG frameworks are not without their own limitations. Most existing RAG systems rely on dense vector embeddings and similarity search over unstructured text corpora. This design assumes a well-defined search space and often requires the number of documents to retrieve to be specified in advance. Such constraints become problematic in scenarios where the search space is unknown, the data is semi-structured or structured (e.g., key-value pairs JSON, tables), or when relationships between entities are complex and multi-faceted. Moreover, traditional RAG pipelines heavily depend on embedding and reranking models to select the most relevant documents, introducing inefficiencies and potential bottlenecks \cite{weller2025theoretical,mendelson2005limitations}.

Graphs offer a compelling alternative for representing and querying complex, interconnected data. By modeling information as nodes and edges, graphs naturally capture relationships, hierarchies, and multi-modal connections that are difficult to express in flat text. Graphs have achieved remarkable success across a wide range of domains, including healthcare, education, recommender systems, and etc. For example, in education, graph-based representations such as knowledge graphs, Bayesian networks and causal graphs facilitate personalized learning pathways and knowledge tracing by capturing dependencies between concepts and student progress \cite{abu2024systematic,donnelly2020impact,yang2015concept,tadayon2021causal}. Recommender systems leverage graph machine learning to model user-item interactions and uncover latent preferences, leading to more accurate and diverse recommendations \cite{wang2021graph,li2013recommendation,gao2022graph}. By representing domain knowledge as directed graphs, these approaches enable powerful reasoning, efficient inference, and the integration of heterogeneous data sources, driving advances in both predictive performance and interpretability.

Graph-based Retrieval-Augmented Generation (Graph RAG) leverages this expressive power, enabling more flexible, context-aware retrieval and reasoning. In \cite{edge2024local}, The authors propose a framework where an LLM constructs a knowledge graph from source documents, detects hierarchical communities within the graph, and generates summaries for each community in a bottom-up manner. For query answering, the system combines relevant community summaries using a map-reduce approach to produce comprehensive answers. In \cite{zhou2025depth}, The authors present a unified framework for Graph RAG, identifying a multiple modular operators that underpin retrieval pipelines. They systematically compare existing Graph RAG methods across diverse QA datasets, introduce new variants that achieve state-of-the-art results, and demonstrate that explicit knowledge graph modeling enhances reasoning, factuality, and interpretability. In \cite{wu2024medical}, The authors present MedGraphRAG, a graph-based Retrieval-Augmented Generation framework tailored for the medical domain to enhance LLM safety and reliability. In \cite{procko2024graph}, the authors provide a survey of approaches that integrate KGs with LLM-based RAG to advance reliable, focused retrieval for specialized applications.

The contributions of this paper are threefold. First, we propose a novel end-to-end Graph RAG framework that constructs RDF-based knowledge graphs from semi-structured JSON data, enabling dynamic and scalable retrieval-augmented generation. Second, we develop a LPG framework by first constructing a text-to-Cypher query generation module, leveraging LLMs to translate user questions into executable Cypher queries in real time by augmenting prompts with graph schema, structure, and metadata. Third, we conduct a comprehensive empirical comparison between our RDF and LPG-based Graph RAG architectures and traditional embedding-based RAG, demonstrating that both graph-based approaches consistently outperform standard RAG in accuracy, reasoning, and response quality. The remainder of this paper is organized as follows: Section 2 reviews background; Section 3 describes the data and experimental setup; Section 4 details our methodology; Section 5 presents results and discussion; Section 6 outlines potential avenues for future research and Section 7 concludes the paper.

\section{Background}
\label{background}
In this section, we provide an overview of some of the key concepts relevant to the rest of paper.
\subsection{Retrieval-Augmented Generation (RAG)}
Retrieval-Augmented Generation (RAG) is a paradigm that enhances large language models (LLMs) by integrating external knowledge sources into the generation process. Unlike standard LLMs, which rely solely on their training data and may produce outdated or hallucinated information, RAG systems dynamically retrieve relevant documents or facts at inference time. This retrieval step grounds the model's responses in up-to-date, authoritative information, significantly improving factual accuracy and reducing hallucinations.

RAG primarily consists of two components: a retriever, which searches a large corpus for passages relevant to the input query, and a generator, which produces responses conditioned on both the query and the retrieved content. As a result, RAG performance is heavily dependent on the quality of the retrieval step, which in turn relies on the effectiveness of the underlying embedding models. Most embedding models are trained on general text and may not be optimized for specialized domains such as finance or healthcare, leading to suboptimal retrieval in these contexts. Furthermore, RAG retrieval is often noisy and highly sensitive to the number of retrieved documents (top-k). To mitigate this noise, reranker models are sometimes employed to reorder the retrieved documents based on relevance before passing them to the generator. However, introducing a reranker increases pipeline complexity and does not necessarily guarantee improved performance.
\subsection{Graph RAG}
Unlike standard RAG, graph-based RAG methods construct a graph from the external corpus to enhance contextual understanding and improve LLM responses. These approaches typically involve three stages: (1) building a knowledge graph by extracting nodes and edges from corpus chunks using an LLM, (2) retrieving relevant nodes or subgraphs from the graph in response to a user query, and (3) generating answers by incorporating the retrieved graph information into the prompt. The key distinctions are that graph-based RAG retrieves and reasons over structured graph elements—such as nodes and relationships—instead of raw text chunks, enabling richer and more precise information retrieval \cite{zhou2025depth}.
\subsection{Resource Description Frameworks (RDF)}
The Resource Description Framework (RDF) is a widely adopted standard for representing information as a graph, particularly in the context of the Semantic Web and knowledge graphs. In RDF, data is modeled as a collection of triplets, each consisting of a subject, predicate, and object. These triplets naturally form a directed, labeled graph where nodes represent entities or values, and edges capture the relationships between them. This structure enables flexible, schema-agnostic modeling of complex, interconnected data and supports powerful querying and reasoning capabilities \cite{pan2009resource,miller1998introduction}.
One can interpret the subject and object in an RDF triplet as nodes, and the predicate as the labeled edge (relationship) connecting them. This mapping allows RDF triplets to be naturally represented as nodes and relationships in a graph structure.
\subsection{Labelled Property Graphs (LPGs)}
The Labeled Property Graph (LPG) is a flexible and widely used graph data model that represents information as nodes, relationships (edges), and properties. In LPG, both nodes and edges can have labels to denote their types, and each can store arbitrary key-value properties, enabling rich, schema-agnostic representation of complex, interconnected data. This structure supports efficient querying, traversal, and reasoning over heterogeneous datasets, making LPG a popular choice for knowledge graphs, social networks, and enterprise data integration \cite{bratanic2024graph}.
\subsection{Graph Query Language (GQL)}
Graph query languages are specialized tools designed to retrieve, traverse, and manipulate data stored in graph structures. For RDF-based graphs, SPARQL is the main querylanguage, enabling expressive pattern matching over triples. SPARQL's design aligns with RDF's semantic foundations, allowing users to extract relationships and infer new knowledge from interconnected data. In contrast, Labeled Property Graphs (LPG) are typically queried using languages such as Gremlin and Cypher (including openCypher). Gremlin is a traversal-based language that enables flexible navigation and analytics over property-rich nodes and edges, while Cypher provides a declarative, SQL-like syntax for pattern matching, filtering, and aggregation. Both Gremlin and Cypher are optimized for LPG's schema-agnostic, property-centric model, supporting efficient graph algorithms and real-time analytics. The choice of query language reflects the underlying graph model: SPARQL for RDF's semantic interoperability, and Gremlin/Cypher for LPG's practical expressiveness and performance \cite{angles2017foundations,francis2018cypher}
\subsection{Amazon Neptune}
Amazon Neptune is a fully managed graph database service designed to support highly connected datasets and complex relationship modeling. Neptune provides native support for both the Resource Description Framework (RDF) and the Labeled Property Graph (LPG) models, enabling users to choose the most suitable paradigm for their application. For RDF data, Neptune implements the SPARQL query language, allowing expressive semantic queries and integration with linked data standards. For LPG, Neptune supports both Gremlin and openCypher, offering flexible traversal and pattern-matching capabilities for property-rich graphs. This dual-model architecture makes Neptune a versatile platform for building knowledge graphs, recommendation engines, and other graph-powered applications, while ensuring compatibility with industry-standard query languages \cite{bebee2018amazon}.
\subsection{NetworkX}
\label{networkx}
NetworkX is an open-source, in-memory graph database and analysis library for Python. It provides a flexible framework for storing and manipulating knowledge graph triplets, allowing users to represent entities and relationships as nodes and edges within a graph structure. NetworkX supports a wide range of graph algorithms and offers powerful visualization tools, making it well-suited for both exploratory analysis and prototyping of graph-based applications. Its ease of use and extensive functionality enable rapid development and experimentation with knowledge graphs, including tasks such as traversal, clustering, and pathfinding \cite{hagberg2008exploring}.
\subsection{FAISS}
\label{faiss}
FAISS (Facebook AI Similarity Search) is an open-source vector database library developed by Meta for efficient similarity search and clustering of dense vectors. It supports both exact and approximate nearest neighbor search, with highly optimized implementations for large-scale datasets. One of its key features is the Hierarchical Navigable Small World (HNSW) algorithm, which enables fast and scalable approximate search with high recall. FAISS is widely recognized for its performance and flexibility, making it a popular choice for embedding-based retrieval tasks in information retrieval, recommendation systems, and large-scale machine learning applications \cite{douze2025faiss}.
\section{Data and Experimental Setup}
\subsection{Dataset}
\label{dataset}
Our experiments utilize publicly available information from Capital Group, focusing on mutual funds, exchange-traded funds (ETFs), and Public Private Solutions (PPS). Each data instance corresponds to a specific investment product, such as the AMCAP Fund (R6 share class), and is structured as a key-value JSON object. In this format, each key represents a fund attribute (e.g., name, product type, benchmark), and the corresponding value provides the attribute’s value. For example, a simplified excerpt for the AMCAP-F3 fund is shown below:

\begin{verbatim}
{
    "originalName": "AMCAP Fund",
    "abbreviatedName": "AMCAP-F3",
    "productType": "AF",
    "benchmarkName": "S&P 500 Index",
    "fundType": "Growth"
}
\end{verbatim}

The full dataset contains 1104 records, each representing a mutual fund, ETF, or PPS product. Notably, many JSON objects exhibit deep nestedness, with certain attributes (such as \texttt{returns} and \texttt{metrics}) containing hierarchical structures up to 3-4 levels deep. This complexity distinguishes the data from simple unstructured text and poses unique challenges for modeling and representation. Since users may query any aspect of the JSON data, it is essential to capture and represent every detail accurately. Appendix~\ref{appendix:nestedness} provides concrete examples of such nested attributes for success metrics. 
\subsection{RDF Data Representation}
\label{RDF_Data_Representation}
To represent our JSON data in the Resource Description Framework (RDF) triplet format, we convert each JSON file—corresponding to a single fund—into a set of subject-predicate-object triplets. We designate the \texttt{abbreviatedName} attribute as the unique subject (source) for all triplets associated with a given fund, as it is both unique and present for every fund in the dataset.

For each key-value pair in the JSON object, we generate a triplet where the subject is the fund's abbreviated name, the predicate is the attribute key, and the object is the attribute value. For example, the sample data shown earlier for the AMCAP-F3 fund is transformed into the following triplets:

\begin{verbatim}
(AMCAP-F3, originalName, AMCAP Fund)
(AMCAP-F3, abbreviatedName, AMCAP-F3)
(AMCAP-F3, productType, AF)
(AMCAP-F3, benchmarkName, S&P 500 Index)
(AMCAP-F3, fundType, Growth)
\end{verbatim}

This process is applied recursively to handle nested attributes, ensuring that all hierarchical levels of the JSON structure are captured as triplets. The approach is both fast and efficient: the entire conversion for all 1,104 funds takes approximately four minutes. Moreover, the method is fully deterministic, introducing no noise or redundancy into the data—unlike LLM-based extraction from unstructured text—thus preserving the integrity and completeness of the original information. Applying this procedure across all funds yields more than 650,000 unique triplets, over 14,000 distinct nodes, and approximately 9,000 unique relationships in the resulting graph.
\subsection{JSON to Text Conversion}
\label{JSON_to_Text_Conversion}
To enable Retrieval-Augmented Generation (RAG) on our structured dataset, the first strategy is to convert each fund's JSON data into unstructured text using a large language model (LLM). For each fund, the JSON object is passed to the LLM with a prompt instructing it to generate a comprehensive narrative that captures all details, regardless of the depth or complexity of nested attributes. This ensures that no information is lost in the conversion process. Along with the JSON, the fund's abbreviated name would be provided in the prompt to help LLM relate the concepts and JSON attributes to the corresponding fund. 

The resulting unstructured text for each fund can then be processed using standard RAG techniques, such as chunking and embedding, to facilitate retrieval and generation. We refer to this representation as RAG\_1. By transforming the structured JSON into text, we can directly compare the performance of traditional RAG pipelines with graph-based approaches on the same underlying data.
\subsection{Embedding Representations}
\label{embedding}
For both the RDF triplet representation (see Section~\ref{RDF_Data_Representation}) and the unstructured text representation (see Section~\ref{JSON_to_Text_Conversion}), we compute embeddings for all unique nodes, relationships, and text segments using the BGE-m3 embedding model. BGE-m3 is a powerful, open-source embedding model selected for its superior performance and versatility \cite{bge-m3}. 

Key properties of BGE-m3 include:
\begin{itemize}
    \item \textbf{Multi-Functionality:} Supports dense retrieval, multi-vector retrieval, and sparse retrieval within a single model.
    \item \textbf{Multi-Linguality:} Capable of processing over 100 languages.
    \item \textbf{Multi-Granularity:} Handles input ranging from short sentences to long documents (up to 8,192 tokens), with an output embedding dimension of 1,024.
\end{itemize}

\subsection{Reranking Operation}
\label{reranking_operation}
For this study, we have utilized several reranking models to evaluate whether accuracy and response quality can be improved in the agentic RAG setting. Reranking operates by taking the initial set of retrieved contexts—based on the user query—and reordering them according to their relevance to the query. This process ensures that the most pertinent information is prioritized for the final response generation, thereby reducing noise and enhancing the overall quality of the retrieved context.

Specifically, we employ two BGE rerankers, \texttt{BAAI/bge-reranker-v2-m3} and \texttt{BAAI/bge-reranker-v2-gemma}~\cite{li2023making}, as well as the widely used \texttt{cross-encoder/ms-marco-MiniLM-L6-v2}. These models leverage powerful transformer architectures to assess the semantic similarity between the user query and each retrieved context segment, assigning higher scores to more relevant passages. By integrating reranking into our pipeline, we aim to systematically investigate its impact on retrieval precision and downstream response generation, particularly in scenarios involving complex, multi-faceted queries.

\subsection{LPG Data Model}
For the development of our data representation layer, we adopt the openCypher query language to model and store JSON data as a Labeled Property Graph (LPG). openCypher is chosen for its expressive power, declarative syntax, and broad adoption in graph database systems, making it well-suited for both data modeling and complex graph analytics.

A significant portion of the design process is dedicated to defining an effective schema—determining which entities should be represented as nodes, refer to ~\ref{dataset}, which relationships should connect these nodes, and which attributes are best modeled as properties. The schema must be carefully crafted to maximize the utility of multihop traversal, a key feature leveraged during inference and query answering. For example, fund entities are modeled as nodes with the \texttt{Fund} label, while attributes such as product type and fund type are represented as separate nodes (\texttt{ProductType}, \texttt{FundType}) and connected to the fund node via explicit relationships (e.g., \texttt{HAS\_PRODUCT\_TYPE}, \texttt{HAS\_FUND\_TYPE}). Other attributes, such as \texttt{originalName} and \texttt{inceptionDate}, are stored as properties on the fund node.

To emphasize the importance of intelligent graph representation, it is noteworthy that our LPG contains approximately 79 unique labels, 57 distinct relationship types, an average of more than 6 properties per label, and over 34,500 nodes. This level of complexity highlights the critical need for careful schema design to ensure efficient storage, retrieval, and scalability in large-scale graph-based systems.

This design enables efficient and flexible graph traversals, allowing queries to exploit the rich connectivity of the data for advanced reasoning and retrieval. An illustrative example of storing a sample JSON object as LPG using openCypher is provided in the Appendix~\ref{lpg_data_model} for reference. 

\subsection{Fund Metadata}
\label{fund_metadata}
We will generate fund metadata based on the original fund data, which consists of information such as fund share classes, fund abbreviation and original name, CUSIP, and product type (e.g., American fund/mutual fund, ETF, PPS/Interval fund, etc.). This metadata is particularly useful in the text2cypher section, as well as for node selection in the RDF pipeline.
\subsection{Clustering the Predicate}
\label{clustering_the_predicate}
Given that, as established in Section~\ref{RDF_Data_Representation}, our dataset contains over 8,000 unique relationships and more than 650,000 unique triplets, and considering that user queries may span multiple categories, intents, and topics, it is essential to categorize relationships based on their semantic similarity. Such categorization streamlines the retrieval process, enabling faster identification of relevant relationships and potentially reducing the number of triplets required to generate a response.

There are two principal approaches to relationship categorization:
\begin{enumerate}
    \item \textbf{Unsupervised Clustering:} In this approach, we compute embeddings for each relationship and apply clustering algorithms, such as K-means or hierarchical clustering, to group similar relationships together.
    \item \textbf{Supervised Classification:} Here, categories are defined in advance based on domain knowledge and comprehensive coverage of key aspects in the data (e.g., fund objectives, metrics, returns). A labeled training dataset is created using a powerful LLM, such as OpenAI GPT-4.1 (the model used in this study), and a model like DeBERTa-v3-large \cite{he2021deberta,he2021debertav3} is fine-tuned to perform large-scale classification. This method significantly reduces inference cost and increases processing speed. An example how this is done is place in appendix~\ref{predicate_clustering_example}.
\end{enumerate}
\subsection{Predicate to Text Generation}
\label{predicate_to_text_generation}
To enhance interpretability and facilitate downstream retrieval, we convert each predicate into natural language. Predicates in our dataset often contain dot (\texttt{.}) notation to denote hierarchical layers, making it essential to accurately reflect this structure in the generated text. Our approach parallels the supervised classification strategy outlined in Section~\ref{clustering_the_predicate}, with the key distinction that we employ the BART Large model instead of DeBERTa-v3. The BART Large model is fine-tuned to translate predicates into natural language descriptions that capture all relevant details and context.

For each predicate, along with its corresponding source and object, we apply deterministic rules to generate clear and comprehensive English sentences. These sentences are then combined, chunked, and embedded for use in the RAG pipeline, a process we refer to as $RAG_2$. An illustrative example of predicate-to-natural-language conversion is provided in the Appendix~\ref{predicate_to_natural_language}.

\subsection{Query Assignment by Intent}
\label{query_assignment_by_intent}
In this section, we describe the categorization of user queries according to business requirements and the nature of the desired response. Queries are divided into four primary intent categories, each reflecting a distinct mode of information retrieval and level of detail:

\begin{enumerate}
    \item \textbf{Compare:} Queries that request comparisons across funds, asset classes, strategies, or other financial entities.
    \item \textbf{Detail:} Queries seeking specific details about a fund or its characteristics.
    \item \textbf{Search/Listing:} Queries that involve listing all funds with a certain property, such as a particular share class or investment strategy.
    \item \textbf{Other:} Queries that do not fit into the above categories and may require special handling or interpretation.
\end{enumerate}

This intent-based categorization guides the response generation process, ensuring that the appropriate level of detail and context is provided for each query type. Representative examples of queries for each intent category are provided in Appendix~\ref{query_assignment_by_intent_example}.

\subsection{Available Fine-Tuned Text2Cypher Models}
\label{available_fine_tuned_text2cypher_models}
At the time of writing, three models fine-tuned for text-to-Cypher translation have been released by the Neo4j team and are available at \url{https://huggingface.co/neo4j}. For our experiments, we specifically evaluated the \texttt{neo4j/text-to-cypher-Gemma-3-27B-Instruct-2025.04.0} model on our use case, which involves a complex dataset with numerous nodes and relationships. The quality of this model was not satisfactory for two main reasons: (1) the model itself lacks sufficient power to handle highly complex datasets, and (2) it consistently selected incorrect relationship and node labels, likely due to its inability to accurately detect the underlying graph topology.

\subsection{Experiment Setup}
\label{experiment_setup}

The following outlines the details of our experimental setup. We sample approximately 200 questions, ranging from medium to very difficult (including some that would challenge human experts), and spanning all four query intents to comprehensively evaluate the quality of our approach. Our evaluation compares the $\mathrm{RAG}_2$ and RDF (triplet-based knowledge graph) representations with the LPG-based representation, measuring response quality in terms of accuracy and completeness. We also discuss the advantages and limitations of each method, drawing general conclusions from our findings beyond the specific context of this experiment.

As explained in Section~\ref{results_discussion}, $\mathrm{RAG}_1$ is excluded from the main experiments due to its lack of scalability for converting JSONs to stories, its limited accuracy, and its susceptibility to hallucination. For further details, please refer to Section~\ref{results_discussion}.
Some examples of queries used in our experiments are provided in Appendix~\ref{query_assignment_by_intent_example}.

\section{Methodology}
\label{methodology}
\subsection{RDF Pipeline: Algorithms}
The RDF pipeline is designed to efficiently retrieve and reason over structured knowledge stored as triplets. As described in Section~\ref{RDF_Data_Representation}, the list of RDF triplets is preprocessed and stored in an Amazon Neptune cluster. In parallel, embeddings for unique nodes and relationships are computed and stored in a vector database (Azure AI is used in this study, though any vector database is suitable).

Unique nodes are extracted from Neptune, where RDF data is stored as subject-predicate-object triplets. These nodes, together with fund metadata (see Section~\ref{fund_metadata}), are passed to a node selection agent. This agent combines LLM-based and deterministic methods: first, an LLM extracts all fund names, abbreviations, fund types, and asset classes mentioned in the user query, using both the query and metadata as input. Next, deterministic mapping is performed to link fund attributes to actual nodes using the available metadata.

The next phase involves identifying relevant relationships based on the user query. The dataset contains over 8,000 distinct relationships, many of which are encoded using dot notation to represent hierarchical structures. Relationships can be complex, often reflecting multi-level hierarchies and nuanced semantics. To address this, we first apply supervised classification as described in Section~\ref{clustering_the_predicate}, which significantly narrows the set of candidate relationships. Subsequently, embedding-based selection is performed to identify up to 50 relevant relationships based on similarity matching. In parallel, the categorized relationships are input to an LLM-based relationship detection agent, which selects up to 100 relevant relationships. The union of unique relationships from both embedding-based and LLM-based methods forms the final set of relationships for retrieval.

Finally, the selected nodes and relationships are used to traverse the Neptune graph using SPARQL, the standard query language for RDF. This traversal returns the set of relevant triplets corresponding to the user query. The same procedure can also be implemented using graph libraries such as NetworkX. The SPARQL output provides a focused set of triplets that serve as the foundation for downstream retrieval-augmented generation.

We replicated the pipeline using NetworkX and FAISS, achieving similar performance. The primary distinction with NetworkX lies in the traversal step, as it operates entirely in-memory and does not utilize SPARQL for graph traversal. In NetworkX, traversal can be implemented by identifying all reachable relationships, intersecting these with the set of selected relationships (determined through a combination of embedding and LLM methods), and retrieving all triplets corresponding to the relevant nodes and relationships.

\begin{figure}[H]
    \centering
    \includegraphics[width=0.85\textwidth]{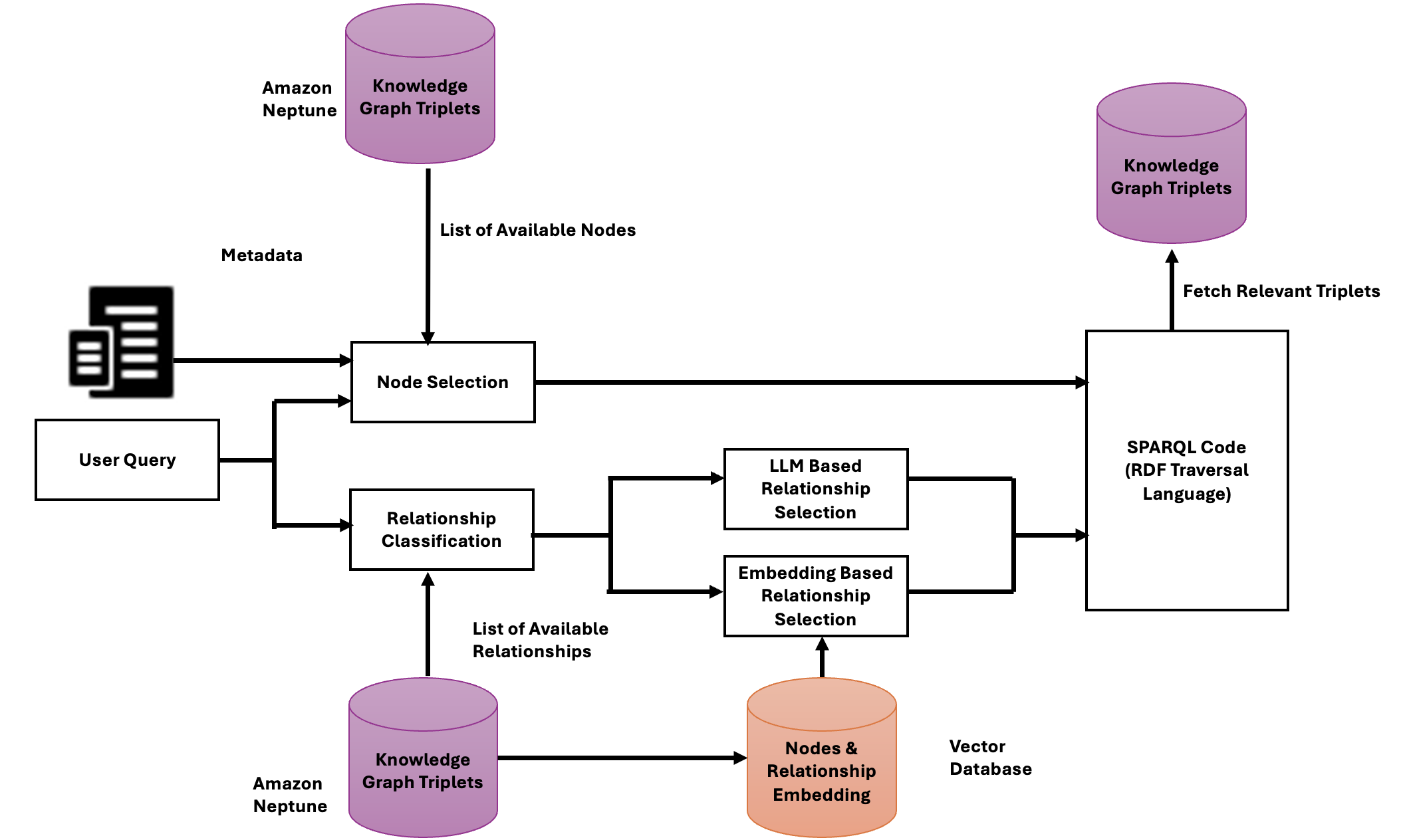}
    \caption{Overview of the RDF-based pipeline for Graph RAG.}
    \label{fig:rdf_pipeline}
\end{figure}

\subsection{LPG Pipeline: Text2Cypher}
\label{lpg_pipeline}

\begin{figure}[H]
    \centering
    \includegraphics[width=0.85\textwidth]{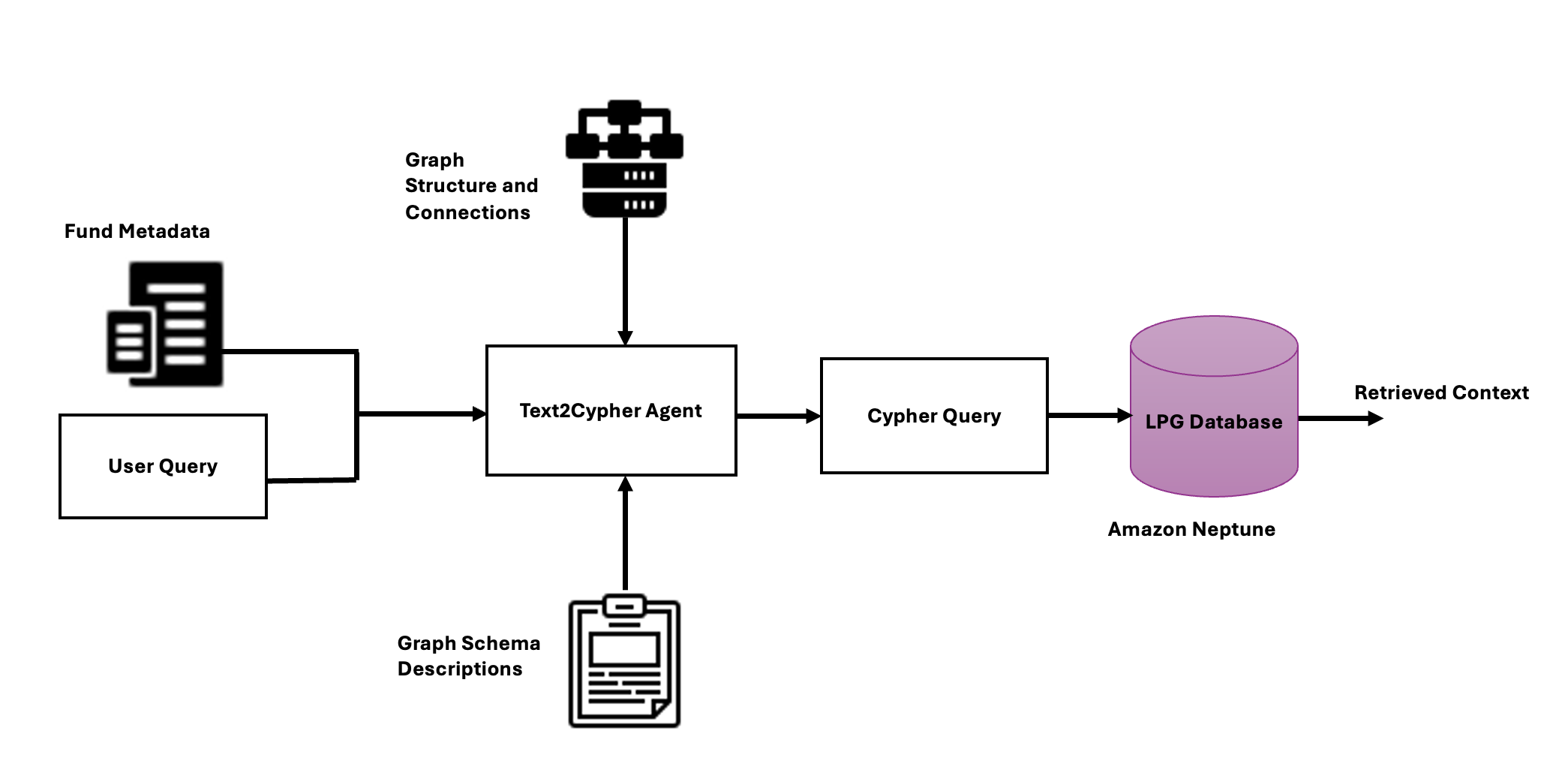}
    \caption{Overview of the LPG-based pipeline for Graph RAG.}
    \label{fig:lpg_pipeline}
\end{figure}

To facilitate accurate text-to-Cypher translation, we generate a comprehensive schema and metadata representation of the Labeled Property Graph (LPG). The schema encodes the graph architecture, including all node labels, properties, and both incoming and outgoing edge types for each node. This structural information enables the text2cypher model to understand how to traverse the graph, which relationships connect nodes, and what properties and labels are available for querying or filtering. Furthermore, we provide detailed descriptions of each node and relationship label, serving as supplementary metadata to inform the model of their semantic roles within the graph.

By providing the model with the full schema, we ensure that generated Cypher queries are both syntactically correct and semantically meaningful, reflecting the true structure of the underlying data. An example Cypher script for extracting all node labels and relationship types from the graph is included in the Appendix~\ref{lpg_schema_query} for reference.
Figure~\ref{fig:lpg_pipeline} provides an overview of this procedure, illustrating the process from user query to context retrieval.
It is important to note that the architectures presented above focus solely on the main representation and retrieval components, and do not include the final response generation, query processing, rewriting, or decomposition steps. For readers interested in these additional stages, an illustrative example is provided in Appendix~\ref{appendix_architecture_example}.
\subsection{Agentic RAG Pipeline}
\label{agentic_rag_pipeline}

In the agentic RAG pipeline, we follow a systematic process to transform structured fund data into a format suitable for retrieval-augmented generation. Initially, the JSON fund data is ingested and converted into RDF triplets, as described in Section~\ref{RDF_Data_Representation}. Each predicate is then translated into natural language using the approach outlined in Section~\ref{predicate_to_text_generation}.

Subsequently, each triplet \texttt{(source, predicate, object)} is converted into a natural language sentence using a rule-based operation of the form: \texttt{source predicate\_text is object}, where \texttt{predicate\_text} is the natural language version of the predicate. All generated sentences for a fund are concatenated to form a comprehensive body of text.

This aggregated text is then segmented into chunks, and embeddings are computed using the BGE-m3 model. The resulting embeddings are stored in a vector database for efficient retrieval. To further enhance retrieval quality, we experiment with various chunk sizes and chunk overlap configurations, as well as different reranker models, to determine the optimal setup for our use case.

\subsection{Evaluation Metrics}
\label{evaluation_metrics}
We evaluate model performance using two key metrics: accuracy and completeness. Accuracy measures whether the provided information is correct and free from hallucination. However, accuracy alone is not sufficient, as a response may be accurate but incomplete or only partially complete. For example, if a user asks a question that requires three reasons and the response includes only two, there is no hallucination, but the answer is incomplete.

Completeness is a more challenging metric to assess, as it requires either a subject matter expert or a well fine-tuned LLM to determine whether all relevant information has been provided. These metrics are highly practical and reflect real-world scenarios, offering a true signal of model performance. In contrast, automated metrics such as ROUGE, BLEU, or BERTScore \cite{papineni2002bleu,lin2004rouge,zhang2019bertscore} do not adequately capture the quality or completeness of responses in retrieval-augmented generation tasks.

\section{Results and Discussion}
\label{results_discussion}

We first address the limitations of $\mathrm{RAG}_1$, which involves converting JSON files into narrative stories using an LLM. As described in Section~\ref{dataset}, the JSON data is highly nested and often very large, making it impractical to fit the entire content into an LLM context window. Even when the data is processed in chunks or batches, several critical issues arise:

\begin{enumerate}
    \item \textbf{Incomplete Information:} Key details are often omitted by the LLM, resulting in stories with missing or incomplete information.
    \item \textbf{Loss of Context:} Context is frequently lost between sections, making it difficult to ensure each attribute is correctly linked to its fund.
    \item \textbf{Hallucination Risk:} The LLM may misrepresent data, especially numeric attributes like returns and statistics, leading to hallucinated responses.
    \item \textbf{Scalability and Cost:} Processing each JSON file, including validation, takes about five minutes. Converting all funds requires roughly four days, which is impractical given frequent data updates.
\end{enumerate}

Due to these scalability and performance challenges, $\mathrm{RAG}_1$ is excluded from further testing and evaluation in this study.
We conduct experiments on a set of 200 questions, as described in Section~\ref{experiment_setup}. Of these, 100 questions pertain to various search and listing tasks, which are typically the most challenging due to their complex search space. Additionally, 45 questions are focused on detail-oriented queries, 45 on comparison tasks, and 10 on other intents. Model performance is evaluated based on response correctness and completeness as described in Section~\ref{evaluation_metrics}. A score of 1 is assigned if the response is both correct and complete; a score of 0.5 is given if the information is accurate but incomplete; and a score of 0 is assigned otherwise.

\begin{table}[H]
\centering
\begin{tabular}{|l|c|}
\hline
\textbf{Method} & \textbf{Score} \\
\hline
RAG$_2$ (Agentic RAG) & 116 \\
$\mathrm{RAG}_{\mathrm{RDF}}$ & 172.5 \\
$\mathrm{RAG}_{\mathrm{LPG}}$ & \textcolor{red}{185.5} \\
\hline
\end{tabular}
\caption{Overall performance scores for each method out of 200 points.}
\label{tab:overall_results}
\end{table}

% \begin{table}[H]
% \centering
% \begin{tabular}{|l|c|c|c|c|}
% \hline
% \textbf{Method} & \textbf{Search} & \textbf{Compare} & \textbf{Detail} & \textbf{Unknown} \\
% \hline
% $\mathrm{RAG}_{\mathrm{RDF}}$ & 0.80 & 0.83 & 0.87 & 0.75 \\
% $\mathrm{RAG}_{\mathrm{LPG}}$ & 0.84 & 0.86 & 0.89 & 0.78 \\
% RAG$_2$ (Agentic RAG) & 0.89 & 0.90 & 0.93 & 0.82 \\
% \hline
% \end{tabular}
% \caption{Performance scores by intent for each method.}
% \label{tab:intent_results}
% \end{table}

\begin{table}[H]
\centering
\begin{tabular}{|l|c|c|c|c|}
\hline
\textbf{Method} & \textbf{Intent} & \textbf{Correct} & \textbf{Partial} & \textbf{Incorrect} \\
\hline
$\mathrm{RAG}_2$ (Agentic)    & Search   & 23 & 31 & 45  \\
$\mathrm{RAG}_2$ (Agentic)    & Compare  & 32  & 6  & 7  \\
$\mathrm{RAG}_2$ (Agentic)    & Detail   & 33  & 3  & 9  \\
$\mathrm{RAG}_2$ (Agentic)    & Other  & 6  & 4  & 0  \\
\hline
$\mathrm{RAG}_{\mathrm{RDF}}$ & Search   & 70 & 20 & 10 \\
$\mathrm{RAG}_{\mathrm{RDF}}$ & Compare  & 37  & 8  & \textcolor{red}{0}  \\
$\mathrm{RAG}_{\mathrm{RDF}}$ & Detail   & 42  & 3  & \textcolor{red}{0}  \\
$\mathrm{RAG}_{\mathrm{RDF}}$ & Other  & 6  & 4  & \textcolor{red}{0}   \\
\hline
$\mathrm{RAG}_{\mathrm{LPG}}$ & Search   & \textcolor{red}{91} & \textcolor{red}{4} & \textcolor{red}{5}  \\
$\mathrm{RAG}_{\mathrm{LPG}}$ & Compare  & \textcolor{red}{39}  & \textcolor{red}{4}  & 2  \\
$\mathrm{RAG}_{\mathrm{LPG}}$ & Detail   & \textcolor{red}{43}  & \textcolor{red}{1}  & 1  \\
$\mathrm{RAG}_{\mathrm{LPG}}$ & Other  & \textcolor{red}{8} & \textcolor{red}{0}  &  2  \\
\hline
\end{tabular}
\caption{Number of correct, partial, and incorrect responses for each method and intent.}
\label{tab:correct_partial_incorrect}
\end{table}

Our experimental results reveal clear performance differences among the LPG, RDF, and agentic RAG approaches across various query intents. For the search/listing intent, the LPG-based method achieves a score of 93 out of 100, substantially outperforming both RDF (80 out of 100) and agentic RAG (38.5 out of 100). In the compare intent, LPG and RDF perform similarly well, with scores of 41 out of 45 each, while agentic RAG lags behind at 35 out of 45. For detail-oriented queries, LPG and RDF again demonstrate strong performance, each scoring 43.5 out of 45, compared to 34.5 out of 45 for agentic RAG. For the other intent, all methods perform comparably, each achieving a score of 8 out of 10.

These results indicate that both LPG and RDF graph-based approaches significantly outperform the agentic RAG pipeline, particularly for search, compare, and detail intents. The superior performance of LPG in the search/listing category can be attributed to its efficient handling of nodes and relationships, as well as its support for multi-hop traversal, which is essential for complex queries involving multiple entities and attributes. 

However, the LPG approach is not without its challenges. The primary source of error arises from confusion or mistakes in converting natural language queries to Cypher, particularly in selecting the correct property or label names for entities such as ETFs, mutual funds, or PPS. These errors can lead to incorrect query results despite the underlying model's strong retrieval capabilities. Overall, the findings underscore the advantages of graph-based retrieval—especially LPG—for knowledge-intensive tasks, while also highlighting areas for further improvement in query translation and schema alignment.
It is important to note that, for agentic RAG, we utilize the rerankers described in Section~\ref{reranking_operation}, evaluating responses both with and without reranking and selecting the best result for final assessment. The primary challenge with agentic RAG—and with approaches that rely heavily on embedding similarity—is the difficulty in managing the retrieval process when the search space is unknown. Specifically, setting the number of documents to retrieve ($K$) is problematic: if $K$ is too small, critical information may be missed; if $K$ is too large, the retrieved context becomes noisy and overly lengthy, leading to irrelevant information and increased risk of incorrect responses. While powerful rerankers can mitigate some of this noise, they are insufficient when $K$ is excessively large.

A second major issue, not limited to search or listing intents, arises from limitations in the embedding models themselves. Embedding models may fail to capture fine-grained distinctions or may confuse similar names and concepts (e.g., mistaking CGCP for CGCB), resulting in the retrieval of incorrect or irrelevant context. These challenges highlight the inherent limitations of agentic RAG pipelines in scenarios with complex or ambiguous search spaces.
For RDF, errors often arise from selecting incorrect nodes and relationships, as these are directly derived from triplets rather than from a carefully designed graph schema. This increases the likelihood of mistakes in mapping queries to the correct graph elements. In contrast, LPG is a graph data model that emphasizes thoughtful schema design. By explicitly defining nodes, labels, and relationships, we can optimize the graph for multi-hop traversal and group similar attributes, thereby reducing the number of nodes and relationships and improving retrieval efficiency.

\begin{figure}[H]
    \centering
    \includegraphics[width=0.7\textwidth]{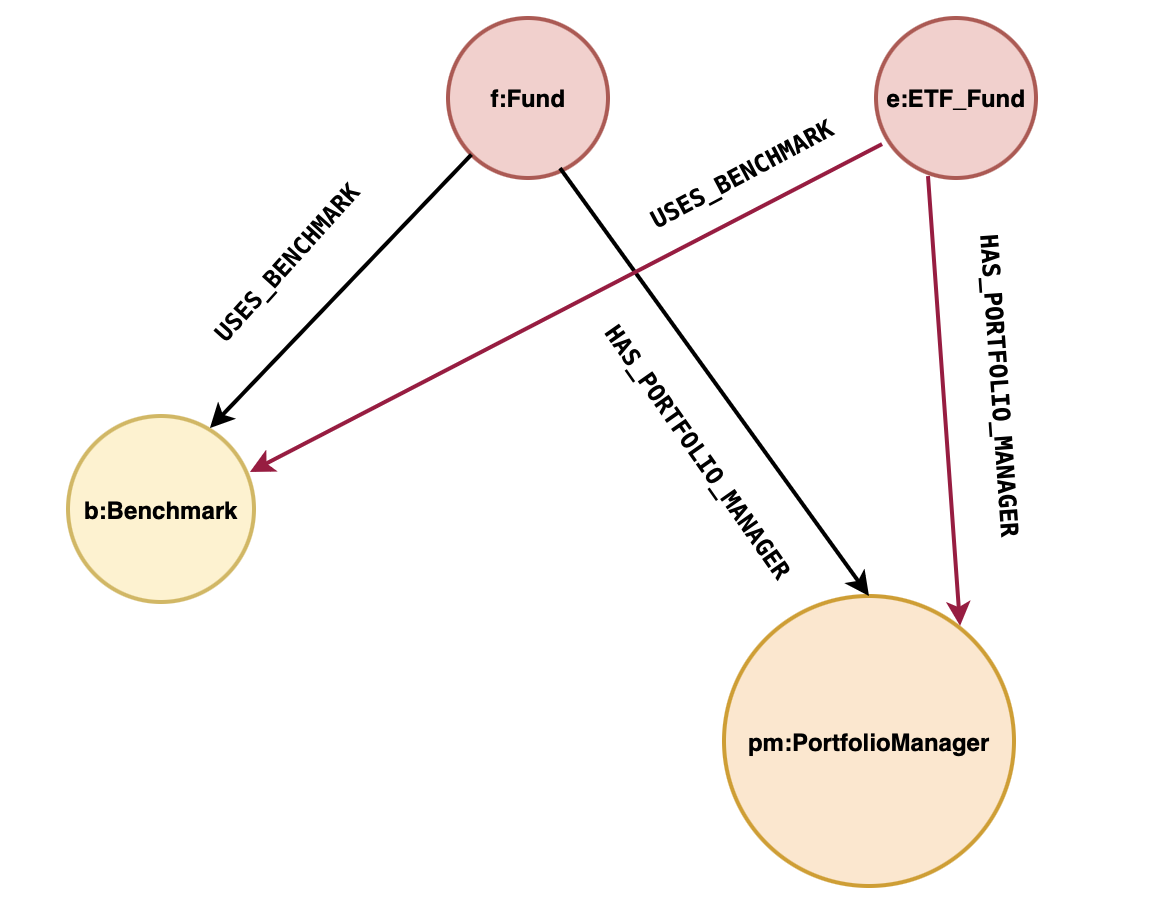}
    \caption{Sample Example of LPG Node Design}
    \label{LPG_nodes_label_design}
\end{figure}

A practical guideline is to create dedicated nodes for each common or important attribute and to reuse these nodes across different fund types (e.g., ETFs, mutual funds, PPS). This normalization of labels and relationships minimizes errors during the text-to-Cypher conversion stage. Figure~\ref{LPG_nodes_label_design} illustrates a sample LPG node design for four labels—benchmark, mutual fund, ETF, and portfolio managers—demonstrating that queries related to portfolio managers or benchmarks can be resolved with a single-hop traversal.

Figure~\ref{LPG_nodes_label_design} illustrates the advantages of LPG schema design for queries such as “List all portfolio managers for AMCAP fund” or “List all funds with S\&P 500 benchmark index.” In the agentic RAG approach, successful retrieval depends on whether all relevant portfolio managers are included in the retrieved context and whether the context is sufficiently noise-free to support reliable response generation. This approach also relies heavily on the embedding model’s ability to accurately capture fund names like “AMCAP”; otherwise, fine-tuning is required, and there is no guarantee of correctness. For more complex queries, such as “List all funds managed by a specific portfolio manager,” agentic RAG faces a fundamental limitation: the number of relevant documents to retrieve is unknown in advance, making it difficult to set retrieval parameters and often requiring iterative tuning per question, which is not scalable.
The RDF-based design offers improvements by maintaining explicit references to fund abbreviations as sources and encoding relationships such as portfolio manager assignments. This structure increases the likelihood of retrieving relevant triplets for queries about funds and their managers. However, for queries like “List all funds managed by a given portfolio manager,” the algorithm may still struggle to select the correct nodes, potentially resulting in incomplete or incorrect answers.

In contrast, the LPG approach enables deterministic and efficient traversal. For example, to answer queries about portfolio managers or benchmarks, the system can directly traverse from fund nodes to the relevant “portfolio manager” or “benchmark” labels and select the desired properties. This design ensures accurate and scalable retrieval for both simple and complex queries, as the graph schema is explicitly constructed to support such multi-hop traversals and attribute groupings.

\section{Future Work}
A promising direction for future research is to implement the proposed system within a multi-agent framework that relies primarily on tool calling, without constructing explicit graph-based or other intermediate representation layers. This approach would enable a direct comparison of its performance against the LPG and RDF frameworks, providing valuable insights into the trade-offs between representation-free and graph-based methodologies.

\section{Conclusion}

In summary, our study demonstrates that both graph-based RAG representations—LPG and RDF—significantly outperform agentic RAG, particularly in scenarios where the search space is unknown and the number of documents to retrieve cannot be predetermined. Our text-to-Cypher model further enhances the capabilities of RDF, while the LPG approach exhibits superior performance when the data is accurately modeled as a graph. The proposed design and pipeline showcase the scalability and effectiveness of graph RAG for large-scale, knowledge-intensive applications.

Our results highlight the reliability of our pipeline in handling semi-structured JSON data with a massive number of keys, values, and deeply nested structures. Overall, these findings underscore the clear advantages of graph-based retrieval—especially LPG—for complex information retrieval tasks, while also identifying opportunities for further improvement in query translation and schema alignment.

\bibliographystyle{unsrtnat}
\bibliography{references}  %%% Uncomment this line and comment out the ``thebibliography'' section below to use the external .bib file (using bibtex) .




% 	\bibitem{keshet2016prediction}
% 	Keshet, Renato, Alina Maor, and George Kour.
% 	\newblock Prediction-Based, Prioritized Market-Share Insight Extraction.
% 	\newblock In {\em Advanced Data Mining and Applications (ADMA), 2016 12th International 
%                       Conference of}, pages 81--94,2016.

% \end{thebibliography}
\section{Appendix}
\subsection{Examples of Nested JSON Attributes}
This part provides an example of nested JSON attributes for success metrics.
\begin{verbatim}
	"successMetrics": {
        "asOfDate": "09/30/2025",
        "byYears": [
            {
                "year": "1YR",
                "successRate": "51.5",
                "avgRollingCompositReturn": "13.36",
                "avgRollingIndexReturn": "11.94",
                "avgRollingOutpacedReturn": "6.91",
                "periodCompositeOutpacedIndex": "355",
                "periodCompositeLaggedIndex": "334",
                "frequency": "quarterly"
            },
            {
                "year": "3YR",
                "successRate": "50.7",
                "avgRollingCompositReturn": "12.12",
                "avgRollingIndexReturn": "10.93",
                "avgRollingOutpacedReturn": "5.38",
                "periodCompositeOutpacedIndex": "337",
                "periodCompositeLaggedIndex": "328",
                "frequency": "quarterly"
            },
            {
                "year": "5YR",
                "successRate": "50.9",
                "avgRollingCompositReturn": "12.23",
                "avgRollingIndexReturn": "10.94",
                "avgRollingOutpacedReturn": "4.77",
                "periodCompositeOutpacedIndex": "326",
                "periodCompositeLaggedIndex": "315",
                "frequency": "quarterly"
            },
            {
                "year": "7YR",
                "successRate": "56.6",
                "avgRollingCompositReturn": "12.30",
                "avgRollingIndexReturn": "10.84",
                "avgRollingOutpacedReturn": "3.86",
                "periodCompositeOutpacedIndex": "349",
                "periodCompositeLaggedIndex": "268",
                "frequency": "quarterly"
            },
            {
                "year": "10YR",
                "successRate": "67.5",
                "avgRollingCompositReturn": "12.66",
                "avgRollingIndexReturn": "10.99",
                "avgRollingOutpacedReturn": "3.12",
                "periodCompositeOutpacedIndex": "392",
                "periodCompositeLaggedIndex": "189",
                "frequency": "quarterly"
            },
            {
                "year": "15YR",
                "successRate": "77.9",
                "avgRollingCompositReturn": "12.62",
                "avgRollingIndexReturn": "11.09",
                "avgRollingOutpacedReturn": "2.33",
                "periodCompositeOutpacedIndex": "406",
                "periodCompositeLaggedIndex": "115",
                "frequency": "quarterly"
            },
            {
                "year": "20YR",
                "successRate": "83.9",
                "avgRollingCompositReturn": "12.46",
                "avgRollingIndexReturn": "11.08",
                "avgRollingOutpacedReturn": "1.72",
                "periodCompositeOutpacedIndex": "387",
                "periodCompositeLaggedIndex": "74",
                "frequency": "quarterly"
            }
        ]
    }
\end{verbatim}
\label{appendix:nestedness}

\subsection{LPG Data Model Example}
\label{lpg_data_model}
\begin{verbatim}
{
    "originalName": "AMCAP Fund",
    "productType": "AF",
    "fundType": "Growth",
    "inceptionDate": "05/01/1967"
}
\end{verbatim}

The corresponding Cypher query is:
\begin{verbatim}
MERGE (f:Fund {fundFamily: $fundFamily})
SET f.name = $name,
    f.originalName = $originalName,
    f.inceptionDate = $inceptionDate

MERGE (pt:ProductType {type: $productType})
MERGE (ft:FundType {type: $fundType})

MERGE (f)-[:HAS_PRODUCT_TYPE]->(pt)
MERGE (f)-[:HAS_FUND_TYPE]->(ft)
\end{verbatim}

\subsection{Example: Cypher Code for Schema Extraction}
\label{lpg_schema_query}
The followings are the cypher code for extracting all the labels and all the properties per label.\\
\textbf{Extract all node labels:}
\begin{verbatim}
MATCH (n) RETURN DISTINCT labels(n) AS node_labels
\end{verbatim}

\textbf{Extract all properties for a given label:}
\begin{verbatim}
MATCH (n:LabelName)
UNWIND keys(n) AS prop
RETURN DISTINCT prop
ORDER BY prop
\end{verbatim}

\subsection{Predicate Clustering Examples}
\label{predicate_clustering_example}
\begin{table}[H]
\centering
\begin{tabular}{|p{10cm}|l|}
\hline
\textbf{Predicate} & \textbf{Category} \\
\hline
topx.topHoldings.values.nameAbbvie.name.asOfDate2025-09-30 & allocation \\
\hline
facts.companiesIssuersDate & generic \\
\hline
geographicBreakdown.regions.nameEurozone.countries.nameGreece.\\values.benchmarkPercent.asOfDate2025-09-30 & allocation, benchmark \\
\hline
\end{tabular}
\caption{Example mapping of predicates to categories.}
\label{tab:predicate_category}
\end{table}

\subsection{Predicate to Natural Language Example}
\label{predicate_to_natural_language}
\vspace{-10pt}
\begin{table}[H]
\centering
\begin{tabular}{l|l}
\toprule
\textbf{Original Predicate} & \textbf{Natural Language Predicate} \\
\midrule
facts.companiesIssuers & Facts About Companies Issuers \\
facts.shareholderAccounts & Facts About Shareholder Accounts \\
descriptions.nameHoldingsOutsideTheUS.description & Description of Holdings Outside the US \\
\bottomrule
\end{tabular}
\caption{Examples of converting predicates to natural language.}
\end{table}

\subsection{Query Assignment by Intent Example}
\label{query_assignment_by_intent_example}
\vspace{-10pt}
\begin{table}[H]
    \centering
    \begin{tabular}{|l|p{10cm}|}
        \hline
        \textbf{Intent} & \textbf{Example Query} \\
        \hline
        Compare & Compare AMCAP-F3 and AMCAP-R6? \\
        Compare & Compare Balanced funds and growth funds in terms of strategy? \\
        \hline
        Detail & What is fund objective and strategy of GFA-R6? \\
        Detail & Who are the portfollio managers for WMIF-R2 fund? \\
        \hline
        Search & List all growth funds? \\
        Search & List the expense ratio for all balanced funds? \\
        \hline
        Other & What is the benefit of moving from balanced funds to growth and income funds? \\
        \hline
    \end{tabular}
    \caption{Representative examples of queries for each intent category.}
\end{table}

\subsection{Full RDF RAG Architecture}
\label{appendix_architecture_example}
\begin{figure}[H]
    \centering
    \includegraphics[width=0.95\textwidth]{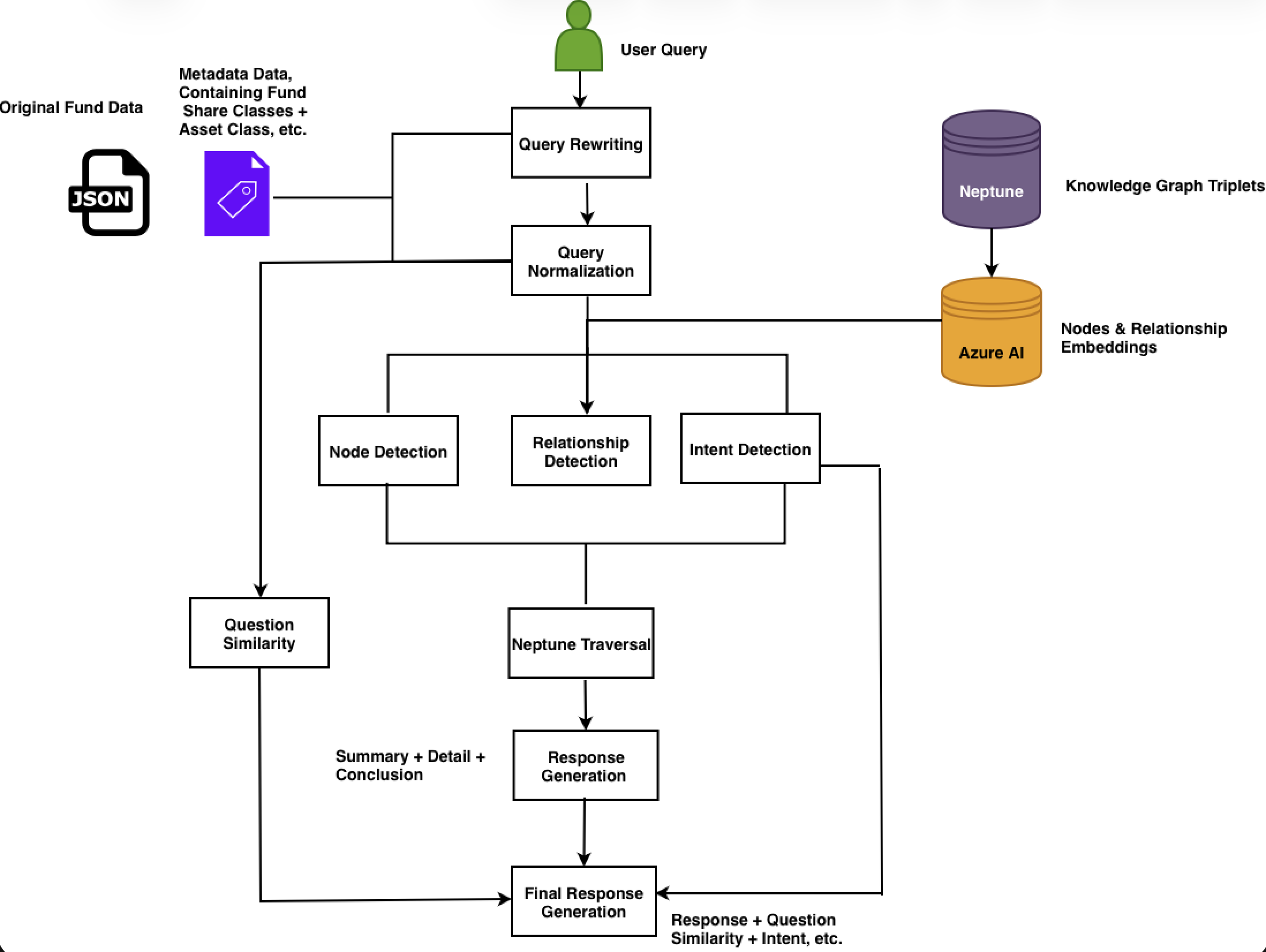}
    \caption{Full Graph RDF structure: This figure illustrates the complete end-to-end pipeline, including all agentic components responsible for query rewriting, normalization, similar query generation, and other orchestration steps.}
    \label{fig:full_rdf_architecture}
\end{figure}

\end{document}